# Modeling Dropwise Condensation underneath Unidirectional Wettability Graded Surfaces

Basant Singh Sikarwar[1*] and Abdelwadood Adil Daoud[2]
[1]Department of Mechanical and Automation Engineering
Amity University Uttar Pradesh
[2] Department of Mechanical and Industrial Engineering
Iowa State University USA
*Corresponding Author: Tel: +91-1204392640, E-mail: bssikarwar@amity.edu

**Abstract**

The dropwise condensation underneath a horizontal super-hydrophobic surface having unidirectional wettability gradient is modeled with implication to enhance the rate of condensation. The mathematical model includes nucleation, growth by vapor condensation and coalescence, and drop instability arising from force imbalances at the three-phase contact line. The wettability graded surfaces, formed by a variable surface energy coating, allow the micro-drop of condensate to slide from the hydrophobic to hydrophilic region without the aid of gravity. The resulting decrease in the drop sliding size shifts the drop size distribution to smaller radii. Furthermore, this decrease in sliding size enhances the heat transfer rate during dropwise condensation. Simulation data has been compared with condensation patterns for a horizontal surface and inclined surface, both with zero wettability gradients. Results obtained show that a wettability graded surface can effectively control the condensation process by decreasing the cycle time of nucleation, growth and removal.

**INTRODUCTION:**

Condensation is a process that occurs in either a dropwise or filmwise mode. Although it has been seen in the day to day life of many individuals, the importance of an understanding of the fundamentals has been overlooked. However, due to the recent push by researchers to understand condensation, its profound effects have been recognized. Specifically, condensation in the dropwise mode has shown the most promising results for enhancing the efficiency of many applications due to its high rate of heat transfer. In energy conversion applications, any method of enhancing the heat transfer coefficient is naturally desirable because of the potential to improve system efficiency which then leads to the conserving of natural resources of energy. Hence, dropwise condensation is an attractive process in a variety of thermal management applications.

In order to enhance the dropwise condensation process, one must understand the hierarchal condensation process entirely. During condensation underneath an inclined substrate, initial nucleation happens at specific sites on the substrate. Leach et al [1] reported a density of approximately $10^4$-$10^6$/cm$^2$ nucleation sites for condensation of water vapor at atmospheric conditions. Once nucleation sites have been established, drops grow by direct condensation of vapor mass on the free surface of the drops. However, when drops reach a certain size, drop growth is mostly attributed to the process of coalescence with neighboring drops. Moreover, although direct condensation is a process that plays a very large role in the initial drop formation, it has been observed to play a much smaller role when the radius of drops grows above 10 micro-meters due to high thermal resistance [2]. When drops research certain critical size, the gravitational force of a droplet exceeds the adhesive force between the droplet and the condensation surface, it depart from the surface. The drop departure process has proven to be an integral step in dropwise condensation as it wipes the surface clean to allow for new nucleation. Once a clean surface has been attained, the process begins again from initial drop growth at specific nucleation sites. This constant cycle ranging from drop formation to drop departure deems dropwise condensation a quasi-cyclic process. Furthermore, research suggests that a higher frequency of the cycle leads to a higher heat transfer rate [3-4]. Thus, any way of increasing the frequency of the quasi-cyclic process may yield very promising results.

Many researchers [5-8] reported that the higher heat transfer coefficient can be attributed to the super-hydrophobicity and vertical orientation of a substrate. These two factors collectively assist to reduce the size of a drop at which the three phase contact line forces are imbalanced due to gravity [8]. Hence, drop size distribution of smaller radii exists on a super-hydrophobic vertical substrate. Therefore, efficient carrying out of the drop mode of condensation depends not only on maintaining hydrophobicity of the substrate but also on controlling the size of mobilized drops.

Due to the intricacies and importance of drop mobilization, researchers [9] have attempted to find simple and effective techniques to influence efficient dropwise condensation. A prime example of a simple approach for mobilization of drops is to incline the substrate with respect to the horizontal; then gravity, surface tension, and pressure are in competition. Force-imbalance is generated by drop deformation and hysteresis, making the apparent advancing and receding angles unequal. Other methods to spontaneously mobilize a drop due to force imbalances on or underneath horizontal surfaces have been reported in the literature [10-14]. These methods consist of: (i) applying a temperature gradient on the hydrophobic substrate so that Marangoni motion is initiated (ii) creating surface energy gradients on the substrate by introducing a suitable physic-chemical treatment (iii) using the aid of magnetism on a horizontal substrate to induce mobilization.

Zamuruyev et al and Yasuda [10-11] experimentally observed the drop motion that results from a patterned surface energy graded hydrophobic surface. Their research promotes the idea that droplets can move on a horizontal surface from the hydrophobic to hydrophilic regions. Schneider et al [12] observed the effects of droplet motion on a horizontal surface with the aid of magnetism. Qiang et al [13] reported that the peak velocity of a water droplet on a horizontal graded surface was 40 mm/s with a distance of movement of 3 mm.

Now, with the advent of nano-technology and breakthroughs in chemical coating technology, the second approach, suitable variable-surface-energy coatings, is possible to move the micro-drops of condensate on/underneath a horizontal substrate without the aid of gravity [14-15]. With this process, droplets are mobilized due to the wettability gradient on the surface which causes a local change in surface energy. Once water drops condense on or underneath the horizontal surface, the drops flow from the hydrophobic region to the hydrophilic region with relatively high speeds [16-18]. Therefore, the gradient in surface energy could be capable of minimizing the time period of the dropwise condensation cycle and thus improving the condensation heat transfer for horizontal surfaces in low gravity environments. Hence, if the substrate orientation is horizontal and limited in some application of dropwise condensation, it is still possible to achieve drop mobilization and a subsequent wiping action by creating a wettability (or variable-surface-energy coating) gradient on the horizontal substrate. Therefore, the spontaneous movement of micro-droplets on a horizontal surface by introducing a suitable wettability gradient is a promising technique to move drops on or underneath the solid surface without external forces. While variable surface-energy coatings have displayed high potential, super-hydrophobicity of the surface will also aid the ability to move condensate drops of smallest radii. In this way, wettability graded super-hydrophobic surfaces would greatly increase heat transfer in dropwise condensation.

Although, the mobilization of a drop due to a wettability gradient on a horizontal substrate is explicitly reported in the literature, very few have applied it to mobilize the drop in dropwise condensation specifically [17]. Daniel et al. [7] experimentally observed more rapid motion of a condensate water drop on a graded horizontal surface. They reported that the heat transfer coefficient of the organic coating and surfactant surface was smaller than the wettability gradient surface. Dietz et al. [2] reported that the passive (due to wettability gradient) removal of drops led to an increase in the heat transfer by a factor of ~14 as compared to filmwise condensation with both surfaces being mounted horizontally. Zhu et al. [18] experimentally observed the self-motion behavior of a condensate drop on a horizontal and an inclined substrate with wettability gradient and reported the velocity of a 2 ml droplet as 42 mm/s on the horizontal and 18 mm/s on the inclined substrate. Their results provide insight into the potential for a horizontal surface that can exhibit efficient dropwise condensation without the aid of gravity.

While condensation underneath a super-hydrophobic wettability gradient surface is efficient, there is no literature which has reported findings from a model of dropwise condensation underneath a super-hydrophobic surface with a wettability gradient. Moreover, the available literature does not give an in-depth analysis of the droplet motion process on wettability gradient which is integral to accurately model dropwise condensation on such a surface. Thus, the process should be focused on and deeply understood for effective modeling of the movements of drops during dropwise condensation. In this background, we present the mathematical model of dropwise condensation underneath a super-hydrophobic horizontal surface having a unidirectional wettability gradient. The model is an extension of the early work of author [20-21], where only inclined substrates without a gradient were considered to be the condensing surface.

The present model includes: initial nucleation, growth by direct condensation and coalescence, and drop instability arising from a surface wettability gradient at the three-phase contact line of drops. The output of the model is: a spatio-temporal drop distribution, area of coverage, wall shear stress, heat transfer rate, available nucleation sites, surface shear stress and the surface heat transfer coefficient. Comparisons of these data are made between inclined and horizontal non-graded surfaces in regards to the potential heat transfer enhancement in dropwise condensation a horizontal substrate with a wettability gradient. The substrate with a wettability gradient allows the smaller drops to slide as compared to a horizontally inclined surface, both having uniform wettability. Therefore, a super-hydrophobic substrate having a unidirectional wettability gradient increases the removal frequency of droplets from the substrate.

**MODEL DEVELOPMENT:**

The model presented here is a variant of an earlier work of the authors [20-21] wherein the dropwise condensation process underneath an inclined, chemical textured (no-graded) substrate was considered. The significant differences here are in modeling of the instability, the terminal velocity and the maximum size of fall-off of a drop underneath a horizontal graded substrate. Hence, first the mathematical model of the mobilization size, terminal velocity and fall-off size of drops are discussed and later the model of the whole dropwise condensation process underneath a horizontal substrate with a wettability gradient will be discussed.

The substrate with wettability gradient, which facilitates droplet motion underneath a horizontal substrate, is shown in Figure 1. Here, the drop is deformed due to variation of wettability on the substrate, Figure 1b. The shape of the contact line is assumed to be circular and the relevant force that moves the liquid phase within the drop towards the hydrophilic region of the substrate is highlighted, Figure 1c.

For calculation of volume and surface area of the deformed drops, their surfaces are fitted by a spherical cap, which is the shape assumed by a small static drops in the absence of a gravitational effect, Figure 1b. The average contact angle of spherical cap approximation droplet is as:

$$\theta_{avg} = \left( \frac{\theta_{max} + \theta_{min}}{2} \right) \tag{1}$$

The drop volume $V$, area of liquid-vapor interface $A_{lv}$ and area of solid-liquid interface $A_{sl}$ are expressed by the following equations:

$$V = \frac{\pi r^3}{3} \left( 2 - 3\cos\theta_{avg} + \cos^3\theta_{avg} \right) \tag{2}$$

$$A_{lv} = 2\pi r^2 \left( 1 - \cos\theta_{avg} \right) \tag{3}$$

$$A_{sl} = \pi r^2 \left( 1 - \cos^2\theta_{avg} \right) \tag{4}$$

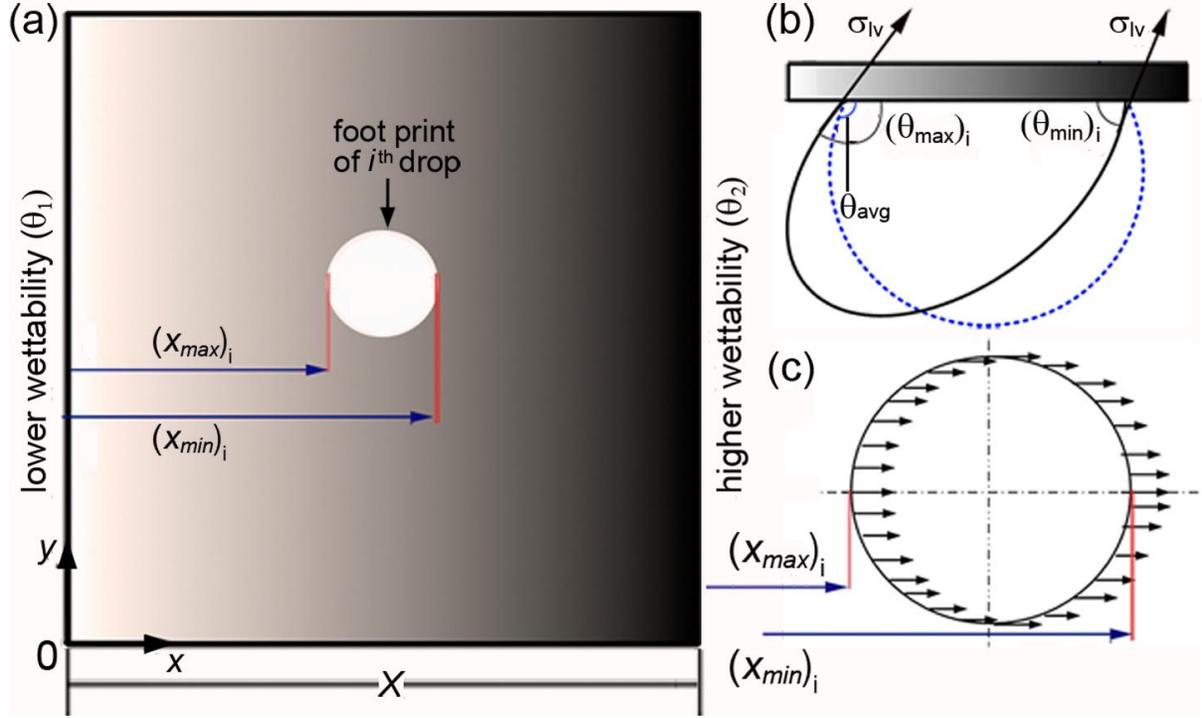

Figure 1: (a) A substrate with wettability gradient. The footprint of $i^{th}$ drop assumed circular. (b) Side view of $i^{th}$ drop and its approximate hemispherical shape (c) direction of force acting at the three phase contact line at substrate with wettability gradient.

The driving force of droplet motion exerted by the substrate on the drop at the contact line due to differences in contact angle (namely, contact angle hysteresis, $\Delta\theta$) is estimated as: For this step, consider $i^{th}$ drop top view of the footprint underneath a unidirectional wettability gradient substrate, assumed to be a circle is used, Figure 2(a-c). The contact angle is taken to vary linearly in one direction from $x = 0$ to $X$, as follows:

$$\theta = \theta_1 - \left(\frac{\theta_1 - \theta_2}{X}\right) \cdot x \tag{5}$$

From Figure 1(c), it can be seen that the net contact line force that acts on the $i^{th}$ drop in the $x$ direction can be calculated as follows:

1. Calculate the base radius of the drop. This is equal to the radius calculated at the previous step plus the effect of the growth rate of the drop (direct condensation plus coalescence). The base radius of droplet of $i^{th}$ is determined as:

$$(r_b)_i = \left[r_i^{old} + \left(\frac{r_i^{new} - r_i^{old}}{\Delta t}\right)\right] \sin(\theta_{avg})_i \tag{6}$$

2. Calculate the $x_{min}$ and $x_{max}$ for the $i^{th}$ drop. After knowing these values, the value of $(\theta_{max})_i$ and $(\theta_{min})_i$ are calculated as:

$$(\theta_{max})_i = \theta_1 - \left(\frac{\theta_1 - \theta_2}{X}\right)(x_{max})_i \tag{7}$$

$$(\theta_{min})_i = \theta_1 - \left(\frac{\theta_1 - \theta_2}{X}\right)(x_{min})_i \tag{8}$$

3. The variation of contact angle at footprint of $i^{th}$ droplet is given by:

$$(\theta_x)_i = (\theta_{max})_i + \frac{\pi - \{(\theta_{min})_i + (\theta_{max})_i\}}{[(x_{min})_i - (x_{max})_i]} x \tag{9}$$

4. The net force acting in the $x$ direction on the $i^{th}$ drop is:

$$(F_x)_i = 2\pi\sigma_{lv}(r_b)_i \int_{(x_{max})_i}^{(x_{min})_i} \cos(\theta_x)_i \, dx \tag{10}$$

Hence

$$(F_x)_i = 2\pi\sigma_{lv}(r_b)_i \frac{X}{(\theta_1 - \theta_2)}\left[\sin(\theta_{max})_i - \sin(\theta_{min})_i\right] \tag{11}$$

The estimation of hydrodynamic force, which acts to oppose the motion of drop, was discussed in detailed in Sikarwar et al. [15]. Here, the hydrodynamic of the $i^{th}$ drop is:

$$(F_{hyd})_i = C_f \left(0.5\rho U_i^2\right)(A_{sl})_i \tag{12}$$

The skin coefficient of friction $C_f$ is taken as:

$$C_f = 58 \, Re^{-0.97} (\theta_{avg})_i^{-1.58} \tag{13}$$

By setting the net force exerted by the solid on the fluid in the horizontal direction $(F_x)_i + (F_{hyd})_i = 0$, a result can be obtained for the terminal speed of each drop in the form:

$$U_i = \left[\frac{0.044 \cdot (F_x)_i (\theta_{avg})_i^{1.58}}{\rho^{0.03} \mu^{0.97} (d_b^{1.03})_i}\right]^{1/1.03} \tag{14}$$

If the weight of the drop is higher than the net retention force normal to the plane due to surface tension, it will fall-off. Hence, the falling criterion for the maximum pendant drop size that is gravitationally stable is:

$$(r_{max})_i = \sqrt{\left[\frac{6\sigma_{lv}\sin^3\theta_{avg}}{\rho g(2-3\cos\theta_{avg}+\cos^3\theta_{avg})}\right]\left(\frac{X}{\theta_1-\theta_2}\right)\left[\cos(\theta_{min})_i-\cos(\theta_{max})_i\right]} \quad (15)$$

*Numerical Methodology:* In the present simulation of dropwise condensation underneath a horizontal substrate with unidirectional wettability graded is chosen as square, 25 × 25 mm² in area. Condensation is initiated by nucleation at 625×10⁴ points over this site corresponding to a nucleation site density of 10⁶ sites per cm². This number is commonly encountered in engineered surfaces used in heat exchangers. The nucleation sites are distributed over the area by using a random number generator function in C++ that returns a matrix containing pseudo random number with a uniform probability density function in the range [0, 1]. The distribution proceeds column-wise till all the sites are occupied. The contact angle at the non-wetting side ($x = 0$) is taken as $\theta_1$ = 158°, while that at the higher wettability side ($x = X$), $\theta_2$ = 151°, is taken for super-hydrophobic substrate. The smallest radius of the droplet located at each nucleation site is given by thermodynamic considerations as:

$$r_{min} = \frac{2\sigma v_l T_w}{H_{lv}[T_{sat}-T_w]} \quad (16)$$

The drop growth rate at each nucleation sites by direct condensation is:

$$\frac{dr}{dt} = \frac{4\Delta T_t}{\rho_l H_{lv}}\left[\frac{\left(1-\frac{r_{min}}{r}\right)}{\frac{2}{h_{int}}+\frac{r}{k_c}}\right]\left[\frac{(1-\cos\theta_{avg})}{(2-3\cos\theta_{avg}+\cos^3\theta_{avg})}\right] \quad (17)$$

Subsequently, they grow with time, the first phase of growth being controlled by direct condensation of vapor. Here, interfacial heat transfer coefficient $h_{int}$ is derived from kinetic theory of gases and is expressed as:

$$h_{int} = \left(\frac{2\hat{\sigma}}{2-\hat{\sigma}}\right)\left(\frac{H_{lv}^2}{T_s v_{lv}}\right)\left(\frac{\bar{M}}{2\pi \bar{R}T_s}\right)^{1/2} \quad (18)$$

When two drops touch each other, they are replaced by a drop of equal volume, placed at their resultant center of mass. The distance between two nucleation sites on the substrate, *i* and *j,* is calculated by following equation:

$$l_{ij} = \sqrt{(x_i - x_j)^2 + (y_i - y_j)^2 + (z_i - z_j)^2} \tag{19}$$

Therefore, the coalescence criterion for the present study is stated as:

$$l_{ij} - (r_i + r_j) < 10^{-3} \tag{20}$$

The drops are allowed to grow by direct condensation as well as coalescence. The active and inactive nucleation sites are verified after the coalescence. Base radius ($r_b$), position of the substrate ($x_{min}$ and $x_{max}$), contact angel hysteresis ($\theta_{max}$ and $\theta_{min}$), terminal velocity and criteria of fall-off for each drops is determined by equation (14). As driving forces overcome hydrodynamic forces, drops move towards the higher wettability side. As the drops slide over the substrate, hidden sites underneath the original drop become active and the nucleation process is repeated. As the drop grows, sliding velocity may be increase or decrease because velocity of sliding depends on the base radius, average contact angle and local position of the drop on substrate. If the weight of the drop is higher than the net retention force normal to the plane due to surface tension, it will fall-off.

*Calculation of surface heat transfer*: Heat transfer during dropwise condensation is calculated by knowing the rate of condensation at free surface of drops at each nucleation site of the substrate. The gaps among nucleation sites are assumed inactive for heat transfer and rate of condensation at each nucleation site is estimates by using a quasi-one dimensional approximation for thermal resistances, including the interfacial and capillary resistance at the vapor-liquid boundary and conduction resistance through drop, as given equation (25). By knowing drop distribution underneath a substrate, the heat transfer ($q$) during the condensation, which depends on the active nucleation sites and rate of growth of radius of drops at each nucleation site, is determined as:

Estimate the active nucleation sites underneath a substrate for given time step. N number of active nucleation sites underneath a substrate at given time step ($\Delta t$). The rate of condensation at each active nucleation site, i.e., $i^{th}$ nucleation site is calculated as:

$$(dm)_i = \left[ \rho \frac{\pi}{3} (2 - 3\cos\theta_{avg} + \cos^3\theta_{avg}) \left( \frac{r_{new}^3 - r_{old}^3}{\Delta t} \right) \right]_i \tag{21}$$

Therefore, the average rate of condensation at time step $\Delta t$ is as:

$$dm = \sum_{i=1}^{i=N} (dm)_i \tag{22}$$

Let $t$ is total time of condensation process. Therefore, the average rate of condensation underneath a substrate is as:

$$(m)_{avg} = \sum_{j=1}^{j=M} \frac{(dm)_j (\Delta t)_j}{t} \qquad (23)$$

here, $M (= t/\Delta t)$ number of time step at given time ($t$) of condensation process. It is now possible to estimate the total heat transfer over an area and hence the average heat flux. Average heat transfer and average heat flux during the dropwise condensation are as:

$$q = (m)_{avg} H_{lv}$$

$$q'' = \frac{(m)_{avg} H_{lv}}{A} \qquad (24)$$

The average heat transfer coefficient during the dropwise condensation is as:

$$\bar{h}_c = \frac{q''}{(T_{sat} - T_{wall})} \qquad (25)$$

*Numerical Algorithm*:

The important steps of the numerical algorithm are listed here: (i) Initialize all variables and input material properties; (ii) randomly distribute the nucleation sites ($10^6$/cm$^2$) on the substrate and place drops of minimum radius at all nucleation sites; (iii) calculate the coordinates of the nucleation site and assign contact angles at each of them; (iv) solve Equation 17 by a 4$^{th}$ order Runge-Kutta method and find the new radius; (v) calculate the intermediate distance between the nucleation sites and check for drop coalescence; (vi) calculate the base radius of drop and estimate contact angle according to equation 6-8; (vii) calculates the sliding velocity by applying the force imbalance at three phase contact line; (vii) again calculate the intermediate distance between the nucleation sites and check for drop coalescence (vii) identify the sites already covered by drops and make them hidden, and simultaneously, search for exposed sites and provide a minimum radius drop on such sites; (vi) check for the critical radius of fall-off; (vii) calculate the rate mass of condensation according to equation 22and 23; (xii) repeat (iii)-(xi) again till stopping criterion is met, i.e., the maximum time considered for condensation. Finally net sum the total condensation is estimated according equation 23. The average heat flux and heat transfer coefficient during the dropwise condensation are calculated according to equation 24 and 25.

**RESULTS AND DISCUSSIONS**

Numerical computations are conducted for nucleation site density of $10^6/cm^2$ on: (i) a substrate with a wettability gradient, (ii) an inclined surface (30°), and (iii) a horizontal substrate; the latter two substrates having no wettability gradients. More details of modeling dropwise condensation process on horizontal and inclined surfaces without wettability gradient are available in [8]. Condensation of water occurs on the underside of the substrate of size 25 mm × 25 mm; drops are taken to be in the pendant mode at all times. For the present simulation, the degree of subcooling is $\Delta T_{sat} = 5°C$ with a saturation temperature of 27°C. Simulations show that the model presented above captures the inherent mechanism of dropwise condensation over a surface with a wettability gradient. The features of the condensation cycle are similar to those of an inclined surface reported in literature [21]. Figure 2 show the temporal-spatial drop distribution underneath a horizontal substrate with wettability gradient. The points of difference for a graded surface are: (i) drops shift towards the higher wettability side, (ii) drops of all sizes are in motion, (iii) larger drops acquire greater velocity and (iv) growth and sliding occur simultaneously.

The model also shows spatial distribution of drops at an instant just before the first drop leaves the surface on a graded substrate. This includes the first slide-off from an inclined substrate and the first fall-off from a horizontal substrate are compared in Figure 3. In view of the motion of drops of every size, there is an exposed virgin area behind every droplet on the graded substrate, as seen in Figure 3a. Hence, the active (exposed) area for a wettability gradient surface is greater than other configurations. The time cycle from initial nucleation to the instant when the drop leaves the surface is also given in Table 1. It is a minimum for the surface with a wettability gradient. For surfaces with wettability gradient, smaller drops move with small velocity and larger drops with large velocity. For an inclined substrate, only the drop that reaches the critical size is set in motion. For the horizontal surface, there is no sliding motion possible; the drop falls off at criticality. For the surface with a wettability gradient, the drop may also fall-off due its weight exceeding surface tension. This factor has been included in the simulation. However, for the range of parameters considered, specifically the size of the substrate, fall-off was not realized.

Figure 4 shows the droplet frequency on the three substrates as a function of drop radius, just before the slide/fall-off criticality is achieved. It is clear that the population of small drops on the graded substrate is larger as compared to the other two. Figure 5 shows the area of coverage with respect to time. It is seen that the area coverage for the graded surface is smaller, making the exposed virgin area larger than the other two surfaces. Consequently, the heat transfer coefficient can be expected to be the greatest for a surface with variable wettability. It is also seen that incipience of droplet slide-off event is at an earlier time instant on the graded surface.

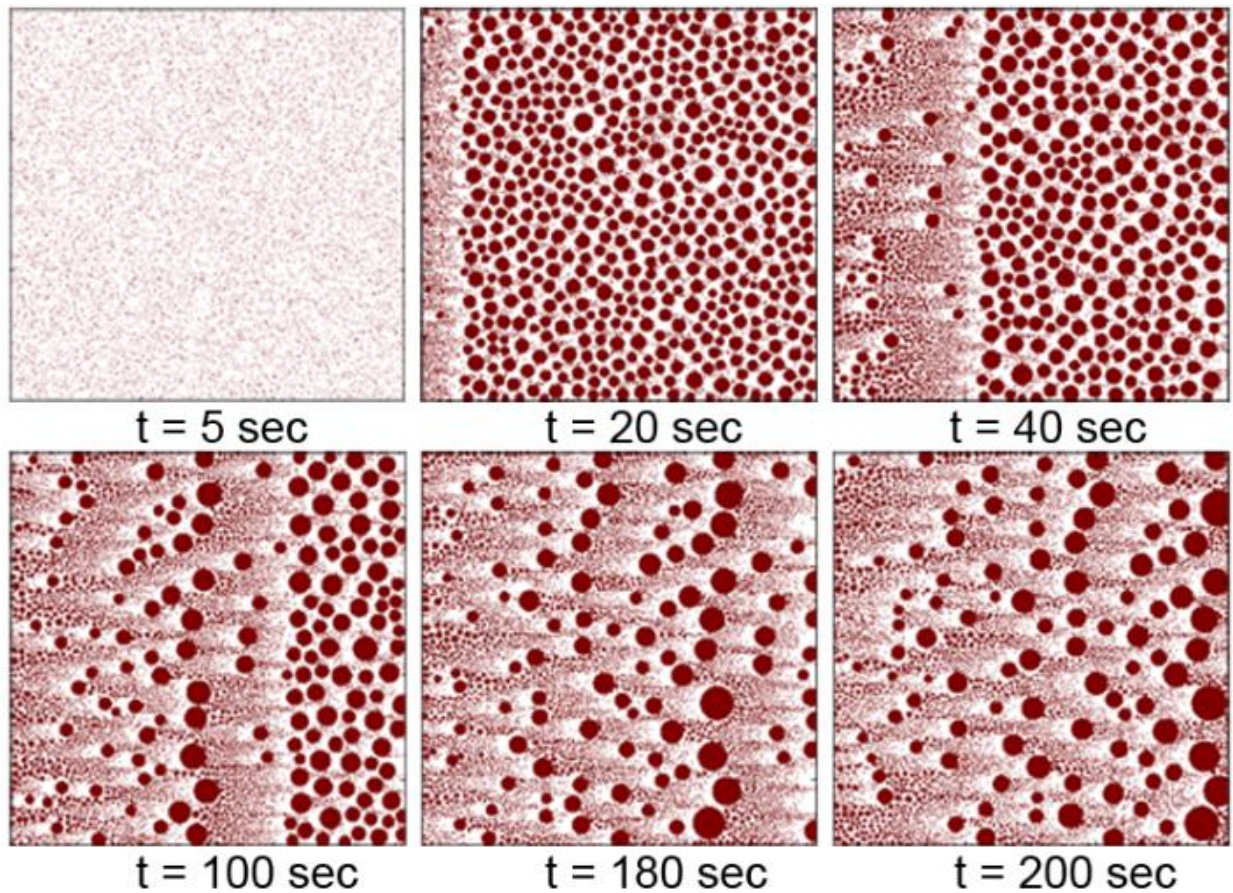

**Figure 2: Temporal-spatial drop distribution underneath a horizontal substrate with wettability gradient.**

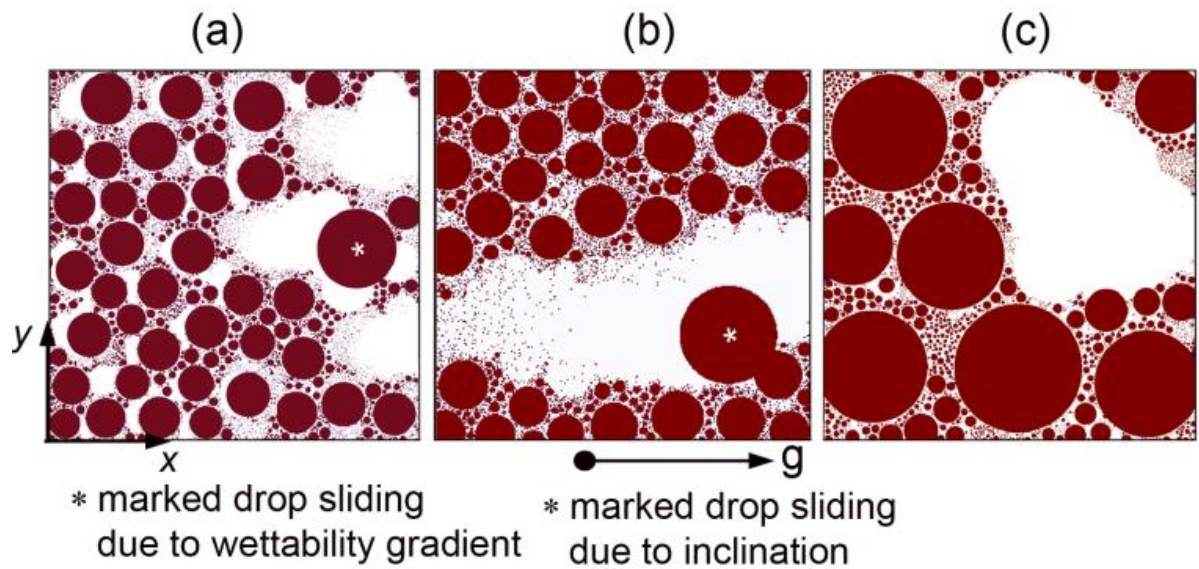

**Figure 3: Droplet motion underneath a chemically textured substrate with (a) wettability gradient imposed on a horizontal substrate, (b) inclined substrate (5°) with no wettability gradient and, (c) a drop falling-off underneath a horizontal substrate with no wettability gradient.**

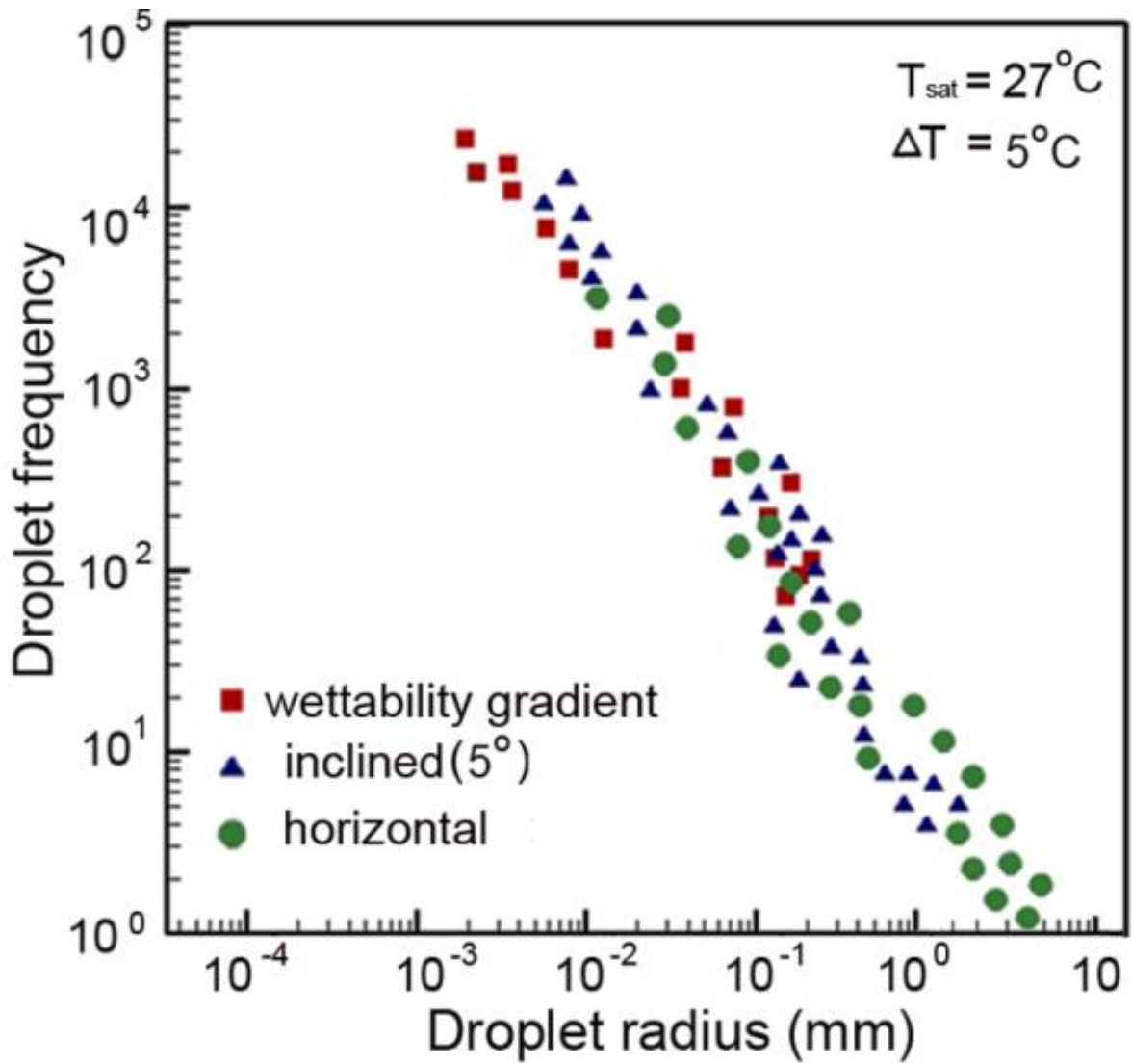

**Figure 4: Droplet frequency (the number of drops) as a function of the drop radius, just before the first drop leaves the surface.**

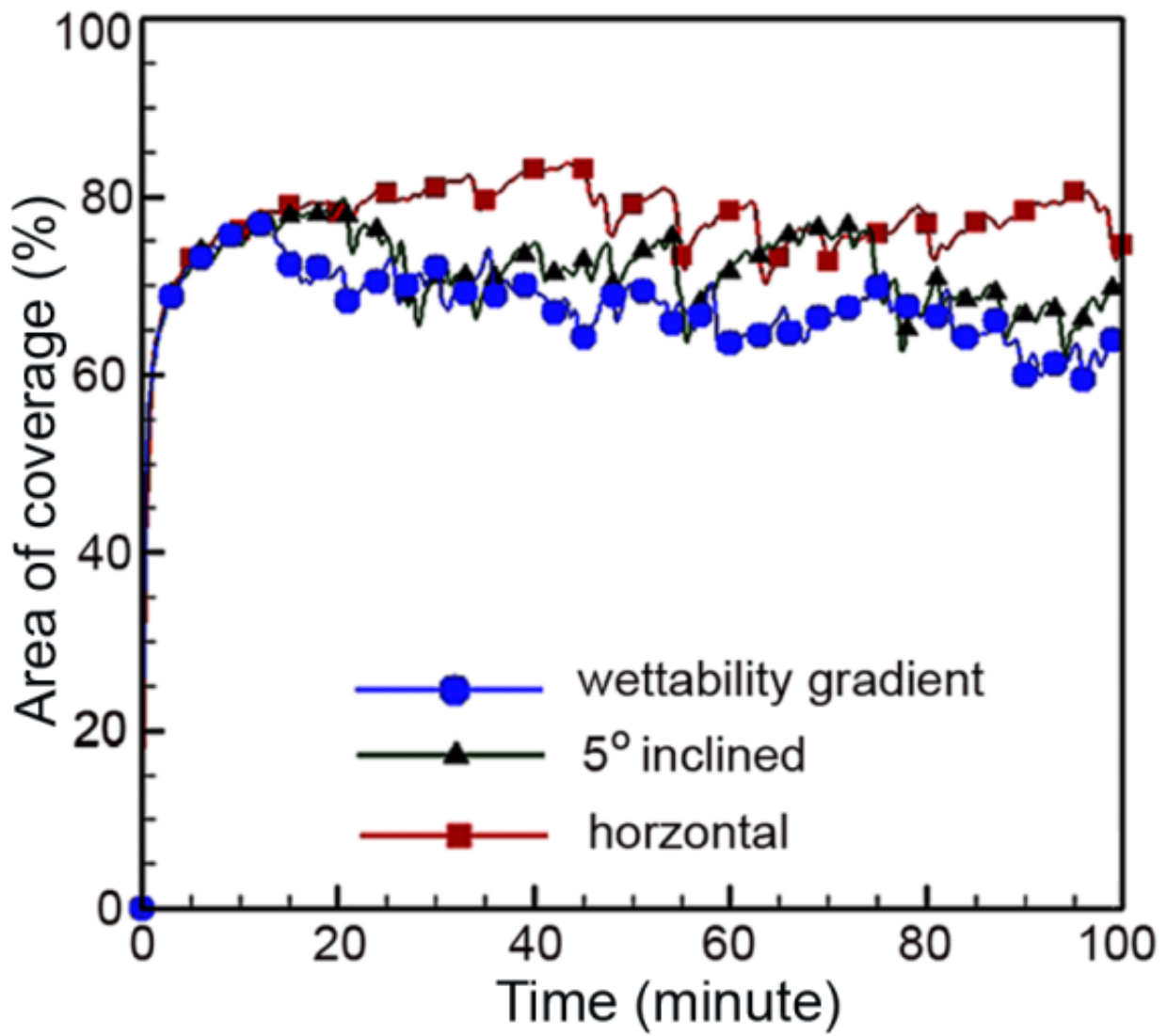

Figure 5: Effect of the choice of the substrate on the percentage area of coverage.

**Table 1: Simulation statistics**

| Case study | Time of first drop slide/fall-off (leave-off) | Size of drop while leaving the surface | Active area |
|---|---|---|---|
| Wettability gradient | 18 minutes | 2.53 mm | 38% |
| Inclined substrate | 42 minutes | 3.47 mm | 23% |
| Horizontal | 55 minutes | 5.67 mm | 18% |

**CONCLUSIONS**

To enhance the efficiency of dropwise condensation, it is crucial to understand the movement of droplets before and after they leave the substrate. To further understand the movement of drops Simulation of dropwise condensation of water vapor underneath a surface having a wettability gradient was carried out. The following conclusions are drawn from the study: It is observed that the velocity of the drop slide depends upon the position of the substrate. Furthermore, the velocity increases as the size of the drop increases. In a wettability gradient substrate, the micro-drop also slides-off toward the higher wettability side. The obtained results show that all sizes of drops move from the hydrophobic to hydrophilic region underneath the horizontal wettability graded condensing surface. Moreover, liquid droplets larger than about 0.5 mm in diameter can move at a peak speed of 150 mm/s with a dependence on the position on the substrate and the mass of droplets. Hence, the self-cleaning process in dropwise condensation due to wettability gradient is a more promising technique for the enhancement of the heat transfer coefficient, as compared to other methods suggested in the literature.

These results possess a very high level of relevance due to the current status of the various applications of dropwise condensation. While research shows that dropwise condensation is ideal, this model paired with current research over dropwise condensation over an energy gradient surface overall allows for dropwise condensation to potentially be implemented into various low gravity applications where efficient condensation is barely applicable. Not only does this data provide insight for dropwise condensation for low gravity situations, but it also exemplifies the power of chemically treated wettability gradients and their effects on the dropwise condensation entirely.

1. Simulation presented herein is sufficient to capture all the major components of the quasi-cyclic dropwise condensation process.
2. Droplets move from a region of lower wettability towards one with higher wettability.
3. On the graded surface, the sliding velocity of drops is a function of its base radius. Larger drops move with higher velocity.
4. The active virgin area for a wettability gradient substrate available for nucleation is greater than what is realized for the other two substrates, i.e. an inclined and a horizontal surface both having uniform wettability.
5. Wettability gradient results in a larger number of small drops and hence will lead to a higher average heat transfer coefficient.

**Acknowledgments:**

This work is an extension of Author (BS) PhD work. So, Author (BS) acknowledges to his PhD advisors Professor K. Muralidhar and Professor Sameer Khandekar for providing simulation facility and guidance for extending the code of dropwise condensation for wettability graded surface. BS is also grateful to his alma matter IITK, India for providing necessary facility to conduct the experiment.